\documentclass[a4paper]{article}
\usepackage{jheppub}
\usepackage[utf8]{inputenc}
\usepackage{amsmath,amsthm,amssymb,mathtools,tikz,booktabs,tensor,slashed,stmaryrd,wrapfig}
\usepackage{todonotes}
\usepackage[mathscr]{eucal}
\newcommand{\bib}[4]{\bibitem{#1} #2, \emph{#3}, #4.} %bibliographic entry: label, names, title, publication details
\newcommand{\biba}[5]{\bibitem{#1} #2, \emph{#3}, #4, \href{https://arxiv.org/abs/#5}{\textcolor[rgb]{0.59, 0.0, 0.09}{#5}}.} %bibliographic entry with arxiv (just put the string that follows after "arxiv.org/abs/")

\newcommand{\mc}{\mathcal} \newcommand{\mf}{\mathfrak} \newcommand{\mb}{\mathbb} \newcommand{\on}{\operatorname}  \newcommand{\ms}{\mathscr} \newcommand{\slot}{\;\cdot\;}
\newcommand{\di}{{\slashed D}}
\title{Supergravity without gravity and its BV formulation}

\author[a]{Julian Kupka,}
\emailAdd{j.kupka@herts.ac.uk}
\author[a]{Charles Strickland-Constable,}
\emailAdd{c.strickland-constable@herts.ac.uk}
\author[a]{and Fridrich Valach}
\emailAdd{f.valach@herts.ac.uk}
\affiliation[a]{Department of Physics, Astronomy and Mathematics,
University of Hertfordshire, College Lane, Hatfield, AL10 9AB, United Kingdom}
\abstract{The generalised-geometric formulation of 10-dimensional supergravity suggests a particular simple ``limit'', which results in a theory whose only dynamical degrees of freedom are the dilaton and the dilatino. The theory is still invariant both under generalised diffeomorphisms and a local supersymmetry and in many aspects is structurally similar to the original supergravity, which makes it a convenient playground for understanding more subtle aspects of the full physical setup. In particular, the simplicity and the geometric nature of the dilatonic theory allow us to build a full BV extension to all orders in the fermionic variables.}

\begin{document}
  \maketitle
\section{Introduction}
  String theory famously features an enormous amount of symmetries and dualities. Part of these is elegantly captured using the framework of generalised geometry \cite{LWX,let,Hitchin,Gualtieri} which, roughly speaking, amounts to the replacement
  \[\text{(tangent bundle, Lie derivative)}\quad \longrightarrow \quad \text{(Courant algebroid, generalised Lie derivative)}.\]
  In this note we focus on performing a particular ``limit'' of the generalised-geometric formulation of the 10-dimensional $\mc N=1$ supergravity coupled to vector multiplets with a gauge group \cite{BRWN,CM,DRSW}. For special choices of the gauge group this is the two-derivative part of the heterotic or type I supergravity. After performing the above-mentioned limit the theory becomes topological, in the sense of not containing any (dynamical) metric degrees of freedom. Before describing the limit procedure, let us stress that this is not related to the twist of supergravity in the sense of Costello--Li \cite{CL}.

  It is known \cite{CSCW,CMTW} that, (at least) up to the quadratic order in fermions, the 10-dimensional supergravity admits a generalised-geometric formulation in which the ordinary fields are naturally interpreted as
  \[\text{metric + $B$-field + gauge field + dilaton}\; \longrightarrow\; \text{generalised metric $\ms G$ + half-density $\sigma$},\]
  \[\text{gravitino + gaugino + dilatino}\; \longrightarrow\; \text{generalised gravitino $\psi$ + generalised dilatino $\rho$}.\]
  The supergravity action can be written, up to quadratic order in fermions, in the elegant form \cite{CSCW}
  \[S_{\text{quad}}=\int \ms R(\ms G,\sigma)\sigma^2-\bar\psi_{\bar a} \slashed D\psi^{\bar a}-\bar \rho \slashed D\rho-2\bar\rho D_{\bar a}\psi^{\bar a}.\]
  Let us comment briefly on the details of the construction. (See the appendix for an introduction to the relevant notions in generalised geometry.) The underlying Courant algebroid here is  transitive, given by the bundle \[E:=TM\oplus T^*M\oplus \on{ad}(G),\] where $\on{ad}(G)$ is the adjoint bundle associated to some principal $G$-bundle, with $G$ a Lie group with an invariant pairing on its Lie algebra. The generalised metric $\ms G$ can be understood as a symmetric endomorphism of $E$ satisfying $\ms G^2=\on{id}$, which induces an eigenbundle decomposition $E=C_+\oplus C_-$. The fields $\rho$ and $\psi$ are an $\mb R$ and $C_-$-valued spinor half-densities w.r.t.\ the subbundle $C_+$.\footnote{These are half-densities w.r.t.\ the base $M$. Note that our formulation with half-densities differs slightly from the treatment in \cite{CSCW}. Although this might seem like a minor point, we will see later that it leads to some drastic simplifications in the formulas, in particular eliminating the need of higher fermionic corrections. We also emphasise here that the way we formulate things here is by using the $O(p,q)$ structure group, as opposed to the equivalent description in \cite{CSCW} in terms of $O(p,q)\times \mb R^+$.} The index $\bar a$ runs over the subbundle $C_-$ and is raised/lowered using the Courant algebroid pairing. In the above physically interesting setup $C_+$ is taken to be the graph of a map $TM\to T^*M\oplus\on{ad}(G)$, which is equi\-valent to the data of $g$, $B$ and $A$. The half-density $\sigma$ is given by $\sigma^2=\sqrt{g}e^{-2\varphi}$, where $\varphi$ is the dilaton understood as a function on $M$. (In this work we will, however, refer to $\sigma$ directly as the dilaton, in accordance with e.g.\ \cite{Siegel}.)
  
  This note is based on the observation that the formalism is consistent even for other (less immediately physically relevant) choices of Courant algebroids and generalised metrics. We will here investigate the most extreme case, namely taking $\ms G$ to be the identity operator, while keeping the Courant algebroid constrained only by the condition that its signature $(p,q)$ is either $(9,1)$ or $(5,5)$\footnote{This condition can be relaxed even more to general Courant algebroids of rank 10, \emph{if} we take all the fields to be complex valued.} and the vector bundle itself is spin (so that we obtain 10-dimensional Majorana--Weyl spinors). This leads to several drastic simplifications. First, the generalised metric becomes non-dynamical, since there is no non-trivial variation of $\ms G=\on{id}$ which is both symmetric and preserves the constraint $\ms G^2=\on{id}$. Furthermore, as the subbundle $C_+$ now spans the entire bundle, there are no generalised gravitinos. The only surviving fields are the dilaton $\sigma$ and the generalised dilatino $\rho$, which is the reason we refer to the resulting theory as the \emph{dilatonic supergravity}.\footnote{This is not related to the other notions of dilatonic supergravity which can be found in the literature.}
  
  The above simplifications allow us to provide a complete generalised-geometric description of the theory to all orders in fermions, together with all its relevant symmetries, and also allow us to write down the full BV action. Our main hope is that some of the insights and structural results can then be carried over to the full physical supergravity.
  
  \begin{wrapfigure}[9]{r}{0pt}
  \begin{tikzpicture}[scale=.7]
    \draw [fill=black] (0,-.95) ellipse (.08cm and .025cm);
    \draw [fill=white] (0,-.7) ellipse (1cm and .25cm);
    \draw [fill=white] (0,-.4) ellipse (1.6cm and .4cm);
    \draw [fill=white] (0,0) ellipse (2cm and .5cm);
    \draw [fill=white] (0,.4) ellipse (1.6cm and .4cm);
    \draw [fill=white] (0,.7) ellipse (1cm and .25cm);
    \draw [fill=black] (0,.82) ellipse (.08cm and .025cm);
    \draw[->,color=blue] (3,.25) -- (2.1,0.1);
    \draw (4.7,.3) node {\scriptsize $g+B$ (exact CA)};
    \draw[->,color=blue] (3,.85) -- (1.6,.6);
    \draw (5.45,.9) node {\scriptsize $g+B+A$ (transitive CA)};
    \draw[->,color=blue] (3,1.4) -- (.2,.86);
    \draw (3.65,1.5) node {\scriptsize $\ms G=\operatorname{id}$};
%    \draw[->,color=blue] (3,-1.2) -- (.2,-1);
%    \draw (3.85,-1.2) node {\scriptsize $\ms G=-\operatorname{id}$};
    \draw[->] (-2.7,-1) -- node[right]{$\!n$} (-2.7,1);
  \end{tikzpicture}
  \caption{Decomposition of $\ms S$ into $\ms S_n$}
  \end{wrapfigure}
  
  We can visualise the passage from the physical supergravity to the dilatonic supergravity as follows. In every fiber of a given Courant algebroid, the space $\ms S$ of possible generalised metrics naturally decomposes into the disjoint union of $\ms S_n$, corresponding to generalised metrics with $\dim C_+=n$. In the transitive case, the physically relevant generalised metrics sit inside $\ms S_{\dim M}$. If $G$ is trivial then $\dim M=\on{rank}(E)/2$, and so the ``physical'' generalised metrics correspond to the largest $\ms S_n$ (in terms of dimensionality). The focus of this paper is on the extremal point $\ms S_{\on{rank}(E)}$ (the top point in the picture), where $\ms G=\on{id}$.
  
  This note is structured as follows: We first describe the dilatonic supergravity and its symmetries. We then provide the full BV construction, count the classical degrees of freedom, and discuss the twisting in the sense of Costello--Li. We look in more detail at two classes of examples, given by exact Courant algebroids and quadratic Lie algebras. Since the latter is particularly simple while still keeping some of the key nontrivial features of the general setup, we have written the relevant Subsection \ref{subsec:qla} in a self-contained way without the Courant algebroid language, in the hope of making it more accessible for the readers without prior acquaintance with generalised geometry. Finally, we provide an introduction to the relevant geometric notions, including Courant algebroids and generalised geometry, in the collection of appendices.

\section{Dilatonic supergravity}
  \subsection{The theory and its symmetries}
  Let $E\to M$ be an arbitrary Courant algebroid with signature either $(9,1)$ or $(5,5)$, with Majorana--Weyl spinor bundles $S_+$ and $S_-$, and denote by $H$ the line bundle of half-densities on $M$.\footnote{For relevant details, definitions, and notation concerning spinors, half-densities, and Courant algebroids, see the Appendices \ref{app:spin}, \ref{app:dens}, \ref{app:Courant}, and \ref{app:connections}. As briefly mentioned before, the existence of $S_\pm$ requires the bundle to be spin, which boils down to certain mild topological conditions, namely the vanishing of the first and second Stiefel--Whitney classes of $E\to M$. In the special case of exact Courant algebroids this is equivalent to the orientability of $M$.}

  The fields of the dilatonic supergravity are the fermionic spinor half-density $\rho\in\Gamma(\Pi S_+\otimes H)$ and the bosonic positive half-density $\sigma\in \Gamma(H)^+$, where $\Pi$ denotes the parity shift. Note that this does not include any metric and hence the theory will be naively topological. The action is
  \begin{equation}\label{eq:action}
    S(\sigma,\rho):=\int_M\ms R\sigma^2-\bar\rho\di\rho,
  \end{equation}
  where $\di$ is the Dirac operator and $\ms R:=\di^2\in C^\infty(M)$.
 The equation of motion for $\rho$ is
 \begin{equation}\label{eq:dirac}
    \di \rho=0,    
 \end{equation}
 while the situation for $\sigma$ is more degenerate. For instance, if $\ms R$ is non-vanishing (almost) everywhere then there are no critical points at all (since $\sigma$ is everywhere positive). On the other hand, if $\ms R=0$ everywhere (as happens for instance on exact Courant algebroids) then the extrema of $S$ are given by any $\sigma$ and $\rho$ satisfying \eqref{eq:dirac}.
  
  In contrast to this almost trivial on-shell structure of the theory, we will now see that the action \eqref{eq:action} still exhibits interesting symmetries analogous to those of the original physical theory, and admits a rich BV extension. For starter, we note that $S$ is invariant under the generalised diffeomorphisms\footnote{Recall that on transitive Courant algebroids these consist of infinitesimal diffeomorphisms, $B$-field gauge transformations, and gauge transformations.} and supersymmetry transformations
  \begin{equation}\label{eq:sym}
      \delta_\zeta \rho=\ms L_\zeta\rho,\quad \delta_\zeta \sigma=\ms L_\zeta\sigma,\quad \zeta\in\Gamma(E),\qquad \delta_\epsilon\rho=\di\epsilon,\quad \delta_\epsilon\sigma=\tfrac1{\sigma}\bar\rho\epsilon,\quad \epsilon\in\Gamma(\Pi S_-\otimes H),
  \end{equation}
  where $\ms L$ is the generalised Lie derivative.
  Note the usefulness of the half-density picture --- replacing the spinor half-densities $\rho$, $\epsilon$ by the more conventional spinors $\rho':=\sigma^{-1}\rho$ and $\epsilon':=\sigma^{-1}\epsilon$, the action and the supersymmetry transformations would take the more complicated form
  \begin{align*}
      S(\sigma,\rho')=\int_M (\ms R-\bar\rho'\di_\sigma\rho')\sigma^2,\qquad 
      \delta_{\epsilon'}\rho'=\di_{\!\sigma}\epsilon'-\tfrac1{96}(\bar\rho'\gamma_{\alpha\beta\gamma}\rho')\gamma^{\alpha\beta\gamma}\epsilon', \qquad \delta_{\epsilon'}\sigma=\sigma(\bar\rho'\epsilon'),
  \end{align*}
  where we defined the ``dressed'' Dirac operator $\di_{\!\sigma}\epsilon':=\sigma^{-1}\di (\sigma\epsilon')$ and used a Fierz identity. Indeed, one of the simplications of using the half-densities is that the Dirac operator $\di$ in the action \eqref{eq:action} is independent of the fields, as opposed to $\di_\sigma$.
  
  Next, a quick calculation with \eqref{eq:sym} gives
  \begin{equation}\label{eq:orig}
      [\delta_{\epsilon_1},\delta_{\epsilon_2}]\rho=0,\qquad [\delta_{\epsilon_1},\delta_{\epsilon_2}]\sigma=\ms L_{\zeta}\sigma,\qquad \zeta^\alpha:=2\sigma^{-2}\bar\epsilon_2 \gamma^\alpha\epsilon_1.
  \end{equation}
  We see that we can (but of course will not) in principle consider the symmetries of the theory to be the local supersymmetry together with the diffeomorphisms which only act on $\sigma$ and leave $\rho$ invariant. However, in order to make connection with the original supergravity and its usual supersymmetry algebra analysis, we will consider the action of diffeomorphisms on $\rho$ as well, and compensate this contribution by another supersymmetry transformation, to obtain a symmetry algebra which closes on-shell. In other words, we can equally well write
  \begin{align}
    [\delta_{\epsilon_1},\delta_{\epsilon_2}]\sigma&=\delta_\zeta\sigma +\delta_\epsilon\sigma\label{eq:vars}\\
    [\delta_{\epsilon_1},\delta_{\epsilon_2}]\rho&=\delta_\zeta\rho +\delta_\epsilon\rho-\tfrac12 \zeta_\alpha \gamma^\alpha \slashed D\rho,\label{eq:varr}
  \end{align}
  where
  \[\zeta^\alpha:=2\sigma^{-2}\bar\epsilon_2\gamma^\alpha\epsilon_1,\qquad \epsilon:=-\tfrac12\zeta_\alpha\gamma^\alpha \rho.\]
  Note that the RHS of \eqref{eq:varr} vanishes due to \eqref{eq:useful}, and the second term on the RHS of \eqref{eq:vars} is zero since $\bar\rho\gamma^\alpha\rho=0$. The rest of the algebra satisfies the usual (off-shell) relations
  \begin{equation}\label{eq:comm}
      [\zeta,\epsilon]=\ms L_\zeta\epsilon,\qquad [\zeta_1,\zeta_2]=\ms L_{\zeta_1} \zeta_2.
  \end{equation}
  The form \eqref{eq:vars}, \eqref{eq:varr}, \eqref{eq:comm} of the local supersymmetry algebra is much simpler than the expressions found in the literature. For instance, in contrast to \cite{BRWN} we only obtain half of the supersymmetric variations (and no Lorentz transformations) on the RHS of \eqref{eq:vars} and \eqref{eq:varr}.
  
  \subsection{BV action}\label{subsec:bv}
    The BV description (which in this case uses $\mb Z_2\times\mb Z$-grading) uses the following fields:
    \begin{center}
    \begin{tabular}{c|c|c|c|c|c|c|c|c|c|c}
      & $\sigma$ & $\rho^a$ & $\xi^\alpha$ & $e_a$ & $f$ & $\sigma^*$ & $\rho_a^*$ & $\xi^*_\alpha$ & $e^{*a}$ & $f^*$\\\hline
      density & half & half & zero & half & zero & half & half & one & half & one\\\hline
      $\mb Z_2$ degree & $[0]$ & $[1]$ & $[0]$ & $[1]$ & $[0]$ & $[0]$ & $[1]$ & $[0]$ & $[1]$ & $[0]$\\\hline
      $\mb Z$ degree & 0 & 0 & 1 & 1 & 2 & $-1$ & $-1$ & $-2$ & $-2$ & $-3$\\\hline
      total & even & odd & odd & even & even & odd & even & even & odd & odd
    \end{tabular}
    \end{center}
    These are all densities valued in either the spinor bundles $S_\pm$, the generalised tangent bundle $E$, the trivial bundle, or their duals (this is marked by the corresponding index which is here Latin for spinors and Greek for sections of $E$), the antifields being decorated by a star.
    The whole BV space can be described as
    \[\ms M_{BV}=T^*[1]\left(\Gamma(H)^+\times\Gamma(\Pi S_+\otimes H)\times \Gamma(E[1])\times\Gamma(\Pi S_-[1]\otimes H)\times C^\infty(M)[2]\right).\]
    The fields $\xi$ and $e$ correspond to the anticommuting ghost for the generalised diffeomorphism symmetry and the commuting ghost for the local supersymmetry, respectively. The field $f$ is a ``ghost for ghost'' corresponding to the fact that $\ms L_{\ms Dh}=0$ for any function $h\in C^\infty(M)$.
    
    The BV action takes the form

    \setlength{\fboxsep}{1pt}
    \begin{center}
    \fbox{\parbox{14.5cm}{
    \begin{align}
      S_{BV}=\int_M \ms R &\sigma^2-\bar\rho\di\rho+\sigma^*(\ms L_\xi \sigma-\sigma^{-1}(\bar \rho e))+\bar\rho^*(\ms L_\xi \rho+\di e)+\bar e^*(\ms L_\xi e+\tfrac12 \sigma^{-2}(\bar e\gamma^\alpha e)\gamma_\alpha\rho) \nonumber\\
      &+\langle\xi^*,\ms D\!f+\tfrac12\ms L_\xi \xi\rangle - \xi_\alpha^*\sigma^{-2}(\bar e\gamma^\alpha e)+\tfrac12 f^* (\ms L_\xi f+ \sigma^{-2}(\bar e\gamma_\alpha e)\xi^\alpha-\tfrac1{6}\langle \xi,\ms L_\xi \xi\rangle)\nonumber\\
      &\vphantom{\int_M} +\tfrac18 \sigma^{-2}(\bar e \gamma_\alpha e)(\bar\rho^*\gamma^\alpha\rho^*).\label{eq:bv}
    \end{align}}}
    \end{center}
    
    Before checking this expression, let us note several interesting features of the action. First, note the appearance of a term of higher order in the antifields --- this corresponds to the fact that our symmetry algebra only closed on-shell. We also note that in the language of $L_\infty$-algebras the last two terms in the second line of $S_{BV}$ correspond to a 3-bracket, schematically
    \[[\text{susy, susy, gauge}]=\text{gauge for gauge},\qquad [\text{gauge, gauge, gauge}]=\text{gauge for gauge}.\]
    
    The verification that \eqref{eq:bv} satisfies the classical master equation $\{S_{BV},S_{BV}\}=0$ is straightforward and will not be shown here in full, since for most part it simply corresponds to the preceding calculation of the supersymmetry algebra. We will however explicitly display below some less trivial parts of the calculation, which also exhibit the general pattern.\footnote{We are using the Koszul sign rule and graded-geometric conventions according to which the Lie derivative satisfies $L_X=[d,i_X]$ (where $[\slot,\slot]$ is the graded commutator), and the Hamiltonian vector field for a function $h$ is defined by $i_{X_h}\omega=-dh$. Since the BV symplectic form $\omega=dp_i\wedge dq^i$ is odd, a (short) calculation shows that for any \emph{even} function $h$ we have \[\{h,\slot\}:=X_h=\frac{\partial h}{\partial p_i}\frac{\partial}{\partial{q^i}}+\frac{\partial h}{\partial q_i}\frac{\partial}{\partial{p^i}},\]
    where we used the \emph{left} derivatives.}
    
    We note that the only nontrivial algebraic identity that is needed in checking the classical master equation is the Fierz identity \eqref{eq:fierz}, and it is due to this identity that we work with spinors in 10 dimensions.
    
    For instance, defining $V\in\Gamma(E)$ by
    \[V^\alpha:=\sigma^{-2}(\bar e\gamma^\alpha e),\]
    the terms in $\tfrac12\{S_{BV},S_{BV}\}$ involving $\rho^*$ and two $e$'s (on top of the ``classical fields'' $\sigma$, $\rho$), combine to give
    \begin{align*}
      \int_M\frac{\delta}{\delta \rho^*}&\left(\int_M \tfrac18 V_\alpha(\bar\rho^*\gamma^\alpha\rho^*)\right)\frac{\delta}{\delta \rho}\left(-\int_M \bar\rho\di\rho\right)+\frac{\delta}{\delta e^*}\left(\int_M \tfrac12 V^\alpha(\bar e^* \gamma_\alpha\rho)\right)\frac{\delta}{\delta e}\left(\int_M \bar\rho^*\di e\right)\\
      & \qquad \qquad +\frac{\delta}{\delta \xi^*}\left(-\int_M V^\alpha\xi_\alpha^*\right)\frac{\delta}{\delta \xi}\left(\int_M \bar\rho^*\ms L_\xi \rho \right) \\
      &=\int_M\tfrac12V_\alpha(\bar\rho^*\gamma^\alpha\di\rho)+\tfrac12V_\alpha(\bar\rho\gamma^\alpha\di\rho^*)-\bar\rho^*\ms L_{V}\rho\\
      &=\int_M\tfrac12\bar\rho V_\alpha\gamma^\alpha\di\rho^*-\tfrac12\bar\rho^*\di V_\alpha\gamma^\alpha\rho=0,
    \end{align*}
    where towards the end we used \eqref{eq:useful}, \eqref{eq:adj}, and the fact that $\overline{\gamma_\alpha\rho}=-\bar\rho\gamma_\alpha$.
    Similarly, using $\gamma_\alpha\gamma^\beta=-\gamma^\beta \gamma_\alpha+2\delta^\beta_\alpha$, the terms involving $\xi^* eee$ give
    \[\int_M-2\sigma^{-2}V^\alpha(\bar \rho e)\xi^*_\alpha+\sigma^{-2}V^\alpha (\bar \rho \gamma_\alpha\gamma^\beta e)\xi^*_\beta=-\int_M \sigma^{-2}V^\alpha (\bar \rho \gamma^\beta \gamma_\alpha e)\xi^*_\beta,\]
    which vanishes due to the Fierz identity \eqref{eq:fierz}. One also easily sees that the last two terms in the second line of $S_{BV}$ are needed to ensure the vanishing of the $\xi^*\xi ee$ and $\xi^*\xi\xi\xi$ terms, respectively.

  \subsection{Degrees of freedom}
    We now briefly examine in more detail the on-shell structure of the theory. We will consider the case of Courant algebroids with $\ms R$ vanishing identically (recall that the other extremal case of everywhere nonvanishing $\ms R$ admits no classical solutions), and count the corresponding degrees of freedom, i.e.\ solutions of the equations of motion modulo gauge transformations. We shall do this on the infinitesimal level, in the following sense.
    
    First, note that a classical background, i.e.\ a bosonic solution to the equations of motion of \eqref{eq:action}, simply corresponds to \[\rho=0,\quad \sigma=\sigma_0,\] with $\sigma_0$ arbitrary. Expanding the BV action around this configuration (we define $\sigma=\sigma_0e^{-\varphi}$), the quadratic part of the action becomes
    \[S_{BV}^{\text{quad}}=\int_M -\bar\rho\di\rho+\sigma^*\ms L_\xi \sigma_0+\bar\rho^*\di e+\langle \xi^*,\ms D\!f\rangle.\]
    From this we read off the linearised BV operator, whose cohomology at degree zero is
    \[H^0=H^0_{\text{even}}\oplus H^0_{\text{odd}},\qquad H^0_{\text{even}}\cong \frac{\{\varphi\in C^\infty(M)\}}{\{\sigma_0^{-1}\ms L_\zeta\sigma_0\mid \zeta\in\Gamma(E)\}},\qquad
    H^0_{\text{odd}}\cong \frac{\{ \di\rho=0\}}{\{\di \epsilon\mid \epsilon\in\Gamma(\Pi S_-\otimes H)\}}.\]
    
    Let us now study in more detail the local degrees of freedom on a transitive Courant algebroid (i.e.\ we will consider a contractible neighbourhood of a point). One can then interpret $H^0_{\text{even}}$ as all functions modulo functions which arise as divergences w.r.t.\ the density $\sigma^2_0$. Since in a contractible region all functions arise in this way, we conclude that $H^0_{\text{even}}=0$. Similarly, one obtains that $H^0_{\text{odd}}$, which coincides with the cohomology of the Dirac operator, is finite-dimensional.\footnote{To see this, we note that on a transitive Courant algebroid over a contractible base the Dirac operator is the sum of the Dirac operators on an exact Courant algebroid and a quadratic Lie algebra --- in the first case this is simply the de Rham differential, while in the second case this is an algebraic operator on the spinors w.r.t.\ the Lie algebra (see Appendix \ref{app:connections}) --- both with finite-dimensional cohomology.}
    Putting things together, we see that $H^0$ is finite-dimensional.
    
  \subsection{Twist \`a la Costello--Li}
    The BV form \eqref{eq:bv} of the dilatonic supergravity provides a simple playground to investigate the twist of supergravity due to Costello--Li \cite{CL}. This amounts to the following question: what are the extrema of $S_{BV}$, or equivalently  at which points in $\ms M_{BV}^{\text{even}}\subset \ms M_{BV}$ does the BV differential $Q_{BV}:=\{S_{BV},\slot\}$ vanish? 
    
    Here $\ms M_{BV}^{\text{even}}$ is the space of configurations with vanishing odd fields, i.e.
    \[\ms M_{BV}^{\text{even}}=\{\rho=\xi=\sigma^*=e^*=f^*=0\}\subset \ms M_{BV}.\]
    Since there are no terms with odd number of odd fields in $S_{BV}$, to find the extrema on $\ms M_{BV}^{\text{even}}$ it suffices to vary the action along the subspace $\ms M_{BV}^{\text{even}}$ itself. Since
    \[S_{BV}|_{\ms M_{BV}^{\text{even}}}=\int_M \ms R \sigma^2+\bar\rho^*\di e - \sigma^{-2}(\bar e\gamma^\alpha e)\xi_\alpha^*+\tfrac18 \sigma^{-2}(\bar e \gamma_\alpha e)(\bar\rho^*\gamma^\alpha\rho^*)+\langle\xi^*,\ms D\!f\rangle,\]
    the extrema correspond to
    \begin{equation}\label{eq:locus}
        \begin{aligned}
            \ms R\sigma^2+(\xi^*_\alpha-\tfrac18 \bar\rho^* \gamma_\alpha\rho^*)\ms D^\alpha\!f=0,\qquad d[a(\xi^*)]=0,\qquad \sigma^{-2}(\bar e\gamma^\alpha e)=\ms D^\alpha\!f,\\\di e=-\tfrac14(\ms D^\alpha\!f)\gamma_\alpha\rho^*,\qquad \di\rho^*=2\sigma^{-2}\xi_\alpha^*\gamma^\alpha e+\tfrac14 \sigma^{-2}(\bar\rho^*\gamma^\alpha\rho^*)\gamma_\alpha e,
        \end{aligned}
    \end{equation}
      where $\ms D^\alpha f=(\ms Df)^\alpha$ and we have used the third equation to simplify the other ones. To understand the second equation, we (locally) pick any orientation and identify\footnote{Recall that $a\colon E\to TM$ is the anchor map (see the appendix).}
      \[a(\xi^*)\in\Gamma(TM\otimes H^2)\cong \Gamma(TM\otimes \Lambda^{\dim M}T^*M)\cong \Omega^{\dim M-1}(M).\]
      
    Costello--Li twisting corresponds to expanding the theory around a background with a nontrivial value of the supersymmetry ghost, i.e.\ around a solution of \eqref{eq:locus} with $e\neq 0$. Note that if we set $f=0$, the supersymmetry ghost must satisfy $\di e=0$ and $\bar e \gamma^\alpha e=0$. Since this corresponds (via \eqref{eq:sym}) to the supersymmetry of the (bosonic) background $(\sigma\in \Gamma(H)^+,\rho=0)$ given by a pure spinor, it can be seen (roughly) as an analogue of the Calabi--Yau condition in the present case. The system \eqref{eq:locus} can then be understood as a generalisation of the Calabi--Yau condition. In particular, the theory obtained by expanding around a solution to this system can be regarded, following the conjecture of \cite{CL}, as an analogue of the BCOV theory \cite{BCOV}. 
      
\section{Examples}\label{sec:examples}
  We will now look in more detail at two classes of examples, lying in a sense at the opposite ends of the spectrum of transitive Courant algebroids --- these are exact Courant algebroids (where the gauge group is trivial) and quadratic Lie algebras (where the manifold is trivial).
  \subsection{Exact Courant algebroids}
    Let $M$ be an oriented 5-dimensional real manifold and $H\in \Omega^3(M)$ a closed 3-form.\footnote{Due to the lack of letters in the alphabet(s), we will be using $H$ for both the 3-form flux and the bundle of half-densities. The distinction should always be clear from the context, and thus the reader will not be confused.} Let $E$ be the corresponding exact Courant algebroid. The bundle of spinor half-densities then corresponds to the bundle of all forms, and chirality translates to the parity of the form degree. In particular, we can choose to identify $S_+\otimes H$ and $S_-\otimes H$ with even and odd forms, respectively, or vice versa. The spinor pairing is given by the Mukai pairing
    \[(\alpha,\beta):=(-1)^{\left[\frac{\deg \alpha}2\right]}(\alpha\wedge\beta)^{\text{top}},\]
    where $(\dots)^{\text{top}}$ extracts the top form part of the expression.
    The Dirac operator is
    \[\di \rho=d\rho+H\wedge \rho.\]
    
    Since for exact Courant algebroids $\ms R=0$, the classical theory is
    \[S(\sigma,\rho)=-\int_M (\rho,d\rho+H\wedge \rho),\]
    where $\sigma$ is a positive bosonic half-density and $\rho$ a fermionic collection of either purely even or purely odd polyforms (depending on which one we pick to correspond to $S_+$ and $S_-$):
    \begin{equation} \label{eq:exact-gauge}
    \sigma\in\Gamma(H)^+,\;\rho\in\Pi\Omega^{\text{even/odd}}(M).
    \end{equation}
    The supersymmetry parameter and transformations are
    \[\epsilon\in\Pi\Omega^{\text{odd/even}}(M),\qquad \delta_\epsilon\rho=d\epsilon+H\wedge \epsilon,\qquad \delta_\epsilon\sigma=\tfrac1\sigma (\rho,\epsilon).\]
    Note that here $\sigma$ naturally decouples both from the action and the transformation of $\rho$. Using integration by parts and taking into account the fermionic nature of $\rho$, we obtain the two theories
    \[\tfrac12S_1=\int_M\rho_0\wedge d\rho_4-\tfrac12\rho_2\wedge d\rho_2+H\wedge\rho_0\wedge \rho_2,\qquad \tfrac12S_2=\int_M\rho_1\wedge d\rho_3-\tfrac12H\wedge\rho_1\wedge \rho_1.\]
    These theories can be regarded as $H$-twisted versions of the $bc$-ghost system.

    In either case we can now consider the BV extension given by \eqref{eq:bv}. We can however also consistently remove all terms containing $\sigma$ or $\sigma^*$, to get
    \begin{align*}
      S_{BV}'=\int_M -(\rho&,\di\rho)+(\rho^*,\di e+\ms L_\xi \rho) + (e^*,\ms L_\xi e) +\langle \xi^*,\tfrac12\ms L_\xi \xi+\ms D\!f\rangle+ f^* (\tfrac12\ms L_\xi f-\tfrac1{12}\langle \xi,\ms L_\xi \xi\rangle).
    \end{align*}
    Note that all the terms are at most linear in antifields, which corresponds to the fact that, with terms with $\sigma$ removed, the supersymmetry closes off-shell. In fact, one can also remove $\xi$, $\xi^*$, $f$, $f^*$ to get simply
    \[S_{BV}''=\int_M -(\rho,\di\rho)+(\rho^*,\di e),\]
    describing a theory acted upon by the supergroup whose Lie superalgebra is purely odd and given by \[\Gamma(\Pi S_-)\cong \Pi\Omega^{\text{odd/even}}(M).\]
    
    Each of $S_{BV}$, $S_{BV}'$, and $S_{BV}''$ thus describe a (different) BV extension of the starting actions $S_1$ and $S_2$. Note that due to $\di^2=0$, the supersymmetry of $S'_{BV}$ and $S''_{BV}$ is (infinitely\footnote{Note that if we consider the action $S_1$ above together with the symmetry transformation~\eqref{eq:exact-gauge}, the gauge parameter $\epsilon$ contains a five-form, which does not generate a variation of $\rho$. This is much like the situation in exceptional generalised geometry \cite{Hull,PW,CSCW2}, where imposing that the gauge parameters assemble into exceptional group representations leads to introducing such parameters. In that situation one can alternatively consider restricting the gauge parameters to forms of those degrees which do generate variations, as one would in supergravity, as implemented in~\cite{CFPSCW}, but this may lead to a different quantum theory. This illustrates that the specification of the gauge transformations should be considered part of the definition of the theory (see e.g.~\cite{AKSZ}).}) reducible. Thus, if we were to quantise these theories, we would need to add a corresponding tower of ghosts to compensate for the reducibility. However, we are not concerned about this in the present text, as our main goal is to gain insight into the symmetry structure of the original physical theory (where this issue does not arise).

  \subsection{Quadratic Lie algebras}\label{subsec:qla}
    One can restrict the analysis of Subsection \ref{subsec:bv} to the particular case when the Courant algebroid is simply a quadratic Lie algebra. However, there is also a way to build a different (but related) BV extension of \eqref{eq:action} for a quadratic Lie algebra, which is what we now turn to. We shall spell out the details, making it self-contained and accessible to readers without prior knowledge of generalised geometry or Courant algebroids.
    
    The setup now corresponds to a given Lie algebra $\mf g$ with invariant pairing of signature either $(9,1)$ or $(5,5)$. Some nontrivial examples include\footnote{Again, if we allow the fields to take complex values, we can drop the signature requirement and only impose $\dim \mf g=10$. Then one can consider other examples, such as $\mf g=\mf{so}(5)$.}
    \begin{itemize}
      \item $\mf g=\mf{su}(2)\oplus \mf{su}(2)\oplus \mf{su}(2)\oplus \mf{u}(1)$ with inner product of signature $(3,0)+(3,0)+(3,0)+(0,1)$
      \item $\mf g=\mf{sl}(2,\mb R)\oplus \mf{su}(2)\oplus \mf{su}(2)\oplus \mf{u}(1)$ with inner product of signature $(2,1)+(3,0)+(3,0)+(1,0)$
      \item the semi-abelian Drinfeld doubles $\mf g=\mf a\ltimes \mf a^*$, for $\mf a$ any 5-dimensional Lie algebra (here $\mf a^*$ is abelian and acted upon by $\mf a$, and $\langle \slot,\slot\rangle$ is the pairing between $\mf a^*$ and $\mf a$).
    \end{itemize}
    Denoting the structure coefficients of $\mf g$ by $c_{\alpha\beta\gamma}$, we have the Dirac operator
    \[\slashed D=-\tfrac1{12}c_{\alpha\beta\gamma}\gamma^{\alpha\beta\gamma}.\] One easily verifies, using the Jacobi identity, that
    \[\ms R:=\slashed D^2=\tfrac12\{\slashed D,\slashed D\}=\tfrac1{144}c^{\alpha\beta\gamma}c_{\delta\epsilon\zeta}(9\delta^\delta_\alpha\gamma_{\beta\gamma}{}^{\epsilon\zeta}-6\delta^{\delta\epsilon\zeta}_{\alpha\beta\gamma})=-\tfrac1{24}c^{\alpha\beta\gamma}c_{\alpha\beta\gamma}\in\mb R.\]
    
    The classical fields of our theory are now $\sigma\in\mb R^+$ and $\rho\in\Pi S_+$, where again $\Pi$ denotes the parity shift and $S_+$ stands for positive chirality Majorana spinors w.r.t.\ the pairing on $\mf g$. The action for the theory is
    \begin{equation}\label{eq:ez}
      S(\sigma,\rho)=\ms R\sigma^2-\bar\rho\di\rho.
    \end{equation}
    This is invariant under the generalised diffeomorphism \eqref{eq:sym}
    \[\delta_\zeta\rho=\tfrac14 \zeta^\alpha c_{\alpha\beta\gamma}\gamma^{\beta\gamma}\rho,\qquad \delta_\zeta\sigma=0,\qquad \zeta\in\mf g,\]
    as well as under the supersymmetry transformations
    \[\delta_\epsilon\rho=\slashed D\epsilon,\qquad \delta_\epsilon \sigma=\sigma^{-1}\bar\rho\epsilon,\qquad \epsilon\in \Pi S_-.\]
    However, we note that in this case we have (cf.\ \eqref{eq:orig})
    \[[\delta_{\epsilon_1},\delta_{\epsilon_2}]=0,\]
    i.e.\ the symmetry algebra closes off-shell into the Lie superalgebra
    \[\mf g\ltimes \Pi S_-,\qquad [\zeta_1+\epsilon_1,\zeta_2+\epsilon_2]=[\zeta_1,\zeta_2]_\mf g+\tfrac14 \zeta^\alpha_1 c_{\alpha\beta\gamma}\gamma^{\beta\gamma} \epsilon_2-\tfrac14 \zeta^\alpha_2 c_{\alpha\beta\gamma}\gamma^{\beta\gamma} \epsilon_1,\]
    which leads to a simple BV description with fields
    \[\sigma\in \mb R^*,\quad\sigma^*\in \Pi \mb R,\quad \rho\in \Pi S_+,\quad \rho^*\in S_-,\quad e\in S_-,\quad e^*\in \Pi S_+,\quad \xi\in\Pi\mf g,\quad \xi^*\in\mf g,\]
    and the BV action
    \begin{align*}
      \tilde S_{BV}=\ms R &\sigma^2-\bar\rho\di\rho+\sigma^{-1}(\bar \rho e)\sigma^*+\bar\rho^*\di e +\tfrac14 \xi^\alpha c_{\alpha\beta\gamma}(\bar\rho^* \gamma^{\beta\gamma} \rho) -\tfrac14 \xi^\alpha c_{\alpha\beta\gamma} (\bar e^*\gamma^{\beta\gamma} e) +\tfrac12 c^\gamma{}_{\alpha\beta}\xi^\alpha\xi^\beta\xi^*_\gamma.
    \end{align*}
    Let us stress again that this is a different BV extension of the same classical theory \eqref{eq:ez} than the one obtained by restricting the analysis in Subsection \ref{subsec:bv} to this case. Similarly, one can consider the action \eqref{eq:bv} with both $f$ and $f^*$ set to zero, i.e.\ restricting to the subspace $\{f=f^*=0\}\subset\ms M_{BV}$. Analogously to the preceding subsection, these three actions define consistent BV theories,\footnote{For the last one this uses the fact that the bracket on $\mf g$ is Lie.} and should be regarded as different BV extensions of the same starting theory \eqref{eq:ez}.\footnote{Again, note that some symmetries might be reducible (e.g.\ if $\mf g$ has a nontrivial centre then there exist $\zeta$'s generating trivial transformations), and so in order to quantise the theory we would need to compensate for this by introducing further ghosts.}
    
    The special property of $\tilde S_{BV}$ is that in checking the classical master equation we do not need to use the Fierz identity \eqref{eq:fierz} and so the theory in fact makes sense on any quadratic Lie algebra with inner product of signature $(p,q)$ with
    \[p+q\equiv 10 \;(\text{mod }8),\qquad p-q\equiv 0 \;(\text{mod }8),\]
    so that we have Majorana--Weyl spinors. Again, taking the fields to be complex-valued we can extend this further to any even-dimensional quadratic Lie algebra. Thus $\tilde S_{BV}$ provides a large class of simple finite-dimensional BV theories which serve as toy models for the original supergravity.
    
\section{Conclusions}
  Although the ultimate goal is to perform a similar analysis in the fully ``physical'' case with (generalised) metrics, the present model already allows us to draw several interesting conclusions that might apply to the structure of the full/physical supergravity.
  
  First, the generalised geometry/Courant algebroid framework tells us that the Dirac operator naturally acts on spinor half-densities \cite{AX,let}. Intending to keep the simple formula from \cite{CSCW} \[\delta_\epsilon\rho=\di \epsilon+\dots,\] in our analysis we were consequently forced to treat both $\epsilon$ and $\rho$ as spinor half-densities. This in turn leads to the following consequences:
    \begin{itemize}
      \item Higher order $\rho$-terms in the supersymmetric variation of $\rho$ drop out. Quite intriguingly, we note that a similar observation was made in \cite{BRWN}, which however uses a different version of dilatino. It is not immediately clear to the authors how these two facts are related.
      \item Half of the supersymmetry variations in the commutator $[\delta_{\epsilon_1},\delta_{\epsilon_2}]$ drop out (see \eqref{eq:vars}, \eqref{eq:varr}). In addition there are no terms with Lorentz transformations.
      \item The Dirac operator $\di$ appearing in the action is independent of the dilaton.
    \end{itemize}
    
    Furthermore, we see that this restricted setup still keeps some nontrivial aspects of the full ``physical'' supergravity story (such as a roughly anticipated form of the BV extension of supergravity), even in the seemingly trivial case of Courant algebroids over a point, i.e.\ quadratic Lie algebras. In fact, if the quadratic Lie algebra $\mf g$ contains a coisotropic subalgebra $\mf h$ then one can canonically construct \cite{LM} a transitive Courant algebroid on the quotient $G/H$ of the corresponding Lie groups --- the bracket on this algebroid will arise from the bracket on $\mf g$. Thus, roughly speaking, nontrivial geometry of $G/H$, reflected in the bracket of its vector fields and hence encoded in the Courant algebroid bracket, corresponds to a nontrivial Lie algebra structure on $\mf g$. This is one of the basic ideas which allow us to extract geometrically interesting results and ideas by working in a purely algebraic framework.

  \acknowledgments

  We are grateful to Pavol \v Severa and Dan Waldram for helpful discussions. C.S.-C.~and F.V.~are supported by an EPSRC New Investigator Award, grant number EP/X014959/1. J.K.\ and F.V.\ would like to thank the organisers of the conference ``Mathematical supergravity'' at the UNED, Madrid, where the idea for this work started. We also thank Miguel Pino Carmona and Tancredi Schettini-Gherardini for the discussion regarding the relevant nomenclature.

  \appendix
  \section{Spinors in signatures \texorpdfstring{$(9,1)$ and $(5,5)$}{(9,1) and (5,5)}}\label{app:spin}
    Let us denote by $S$ the space of Majorana spinors. We will adhere to the conventions (cf.\ \cite{CSCW})
    \[\{\gamma_\alpha,\gamma^\beta\}=2\delta^\beta_\alpha,\qquad \bar\psi:=\psi^TC,\qquad C\gamma_\alpha C^{-1}=-\gamma_\alpha^T,\qquad C^T=-C,\qquad \gamma^{\alpha\dots\beta}:=\gamma^{[\alpha}\dots\gamma^{\beta]}.\]
    which in particular imply that for any pair of fermionic spinors
    \[\overline\psi\gamma_{(p)}\chi=(-1)^{\left[\frac{p+1}2\right]}\overline\chi\gamma_{(p)}\psi,\]
    where $\gamma_{(p)}$ denotes any $\gamma_{\alpha\dots\beta}$ with $p$ indices.
    
    Majorana spinors admit the decomposition $S=S_+\oplus S_-$ into Weyl spinors, where
    \[\gamma^{(10)}|_{S_\pm}=\pm\on{id}_{S_\pm},\qquad \gamma^{(10)}:=\gamma^0\dots\gamma^9.\]
    We also have the following vanishing bilinears:
    \[\overline\psi\gamma_{(2k)}\chi=0 \text{ if $\psi$, $\chi$ have the same chirality},\qquad \overline\psi\gamma_{(2k+1)}\chi=0 \text{ if $\psi$, $\chi$ have opposite chirality.}\]
    This in particular implies that the only nonzero bilinear of a single chiral fermionic spinor $\rho$ is
    \[\bar\rho\gamma_{\alpha\beta\gamma}\rho.\]
    Two further important things that happen in 10 dimensions (which do not happen in $10+8k$ dimensions for $k\ge 1$) are:
    \begin{itemize}
      \item the fourth antisymmetric tensor power of $S_+$ (or $S_-$) does not contain any singlet, implying that one cannot add any quartic terms in the chiral fermionic spinor $\rho$ to the action \eqref{eq:action}
      \item the third symmetric power of $S_-$ does not contain any $S_+$ summand, and vice-versa, implying the following Fierz identity for any bosonic chiral spinor $e$:
    \begin{equation}\label{eq:fierz}
      (\bar e \gamma_\alpha e)\bar e\gamma^\alpha=0.
    \end{equation}
    \end{itemize}
    The latter is the reason why we restrict our analysis to 10 dimensions, since \eqref{eq:fierz} is needed in order that our BV action \eqref{eq:bv} satisfies the classical master equation.
    
  \section{On \texorpdfstring{$\lambda$}{lambda}-densities}\label{app:dens}
    Let $M$ be an $n$-dimensional manifold and $\lambda$ a real number. We define the line bundle $L^\lambda$ as the bundle associated to the frame bundle of $M$ via the 1-dimensional representation of $GL(n,\mb R)$ given by
    \[A\mapsto \left|\det(A)\right|^{-\lambda}.\]
    Practically, this means that every local frame on $M$ induces a local section of $L^\lambda$, and changing the frame by a transition matrix $A$ results in multiplying the section by $\left|\det(A)\right|^{-\lambda}$. Sections of $L^\lambda$ are called \emph{$\lambda$-densities}. If $\lambda=1/2$, we simply talk about \emph{half-densities}; for simplicity we will set \[H:=L^{1/2}.\] We also have $L^{\lambda}\otimes L^{\lambda'}\cong L^{\lambda\otimes\lambda'}$.
    Owing to the absolute value in their definition, $\lambda$-densities enjoy two important properties:
    \begin{itemize}
      \item they always exist globally, i.e.\ the line bundle $L^\lambda$ is always trivial (though it does not have a canonical trivialisation),
      \item it makes sense to talk about positive or negative $\lambda$-densities (at every point on $M$).
    \end{itemize}
    The space of everywhere positive half-densities will be denoted by $\Gamma(H)^+$.
    
    Finally, 1-densities can be naturally integrated. In fact, if $M$ is orientable, a choice of orientation on $M$ provides an identification of $1$-densities with top forms on $M$. However, integration of 1-densities is well-defined even on non-orientable manifolds.
    
  \section{On Courant algebroids}
  \label{app:Courant}
    A \emph{Courant algebroid} \cite{LWX} is a vector bundle $E\to M$, equipped with some additional structure, namely
    \begin{itemize}
      \item an $\mb R$-bilinear operation $[\slot,\slot]\colon \Gamma(E)\times\Gamma(E)\to\Gamma(E)$
      \item a fiberwise non-degenerate bilinear symmetric form $\langle\slot,\slot\rangle$
      \item a vector bundle map $a\colon E\to TM$
    \end{itemize}
    satisfying several axioms.
    First, for all $u,v\in\Gamma(E)$, $f\in C^\infty(M)$ we have \[[u,fv]=f[u,v]+(a(u)f)v.\] This allows us to extend the action of $u$ on sections of $E$ (via $[u,\slot]$) and on functions (via $a(u)$) to a derivation $\ms L_u$ of the whole tensor algebra on $E$. For instance, for $v,w\in\Gamma(E)$ we have \[\ms L_u(v\otimes w)=(\ms L_u v)\otimes w+v\otimes(\ms L_u w).\] We call $\ms L$ the \emph{generalised Lie derivative}. The remaining axioms of the Courant algebroid can then be simply stated as
    \[\ms L_u[v,w]=[\ms L_u v,w]+[v,\ms L_u w],\qquad \ms L_u\langle\slot,\slot\rangle=0,\qquad [u,v]+[v,u]=\ms D\langle u,v\rangle,\]
    for any $u,v,w\in\Gamma(E)$ and $f\in C^\infty(M)$, where the operator $\ms D\colon C^\infty(M)\to\Gamma(E)$ is defined by
    \[\langle \ms Df,u\rangle:=a(u)f.\]
    Note that the pairing/inner product $\langle \slot,\slot\rangle$ allows us to identify $E\cong E^*$, which we will use freely. In other words, the indices on a Courant algebroid are always lowered/raised using this inner product.
    
    The compatibility of $\ms L$ and the pairing imply that $\ms L_u$ also acts naturally on any associated spinor bundles. Finally, we have a natural action on $\lambda$-densities, given by
    $\ms L_u\sigma=L_{a(u)}\sigma$, where $L$ is the ordinary Lie derivative.
  
    Several things can be worked out following this definition. First, for any two sections we get
    \[a([u,v])=[a(u),a(v)],\]
    which implies that if $\on{rank}(a)$ is constant (we say the algebroid is \emph{regular}) then $\on{im}(a)\subset TM$ is an integrable distribution. Furthermore, denoting the dual map to $a$ by $a^*$, we have the important property $a\circ a^*=0$, which can be restated by saying that
    \begin{equation}\label{eq:exact}
      0\to T^*M\xrightarrow{a^*}E\xrightarrow{a} TM\to 0
    \end{equation}
    is a chain complex. We say that a Courant algebroid is \emph{exact} if \eqref{eq:exact} is an exact sequence. More generally, it is \emph{transitive} if $a$ is surjective. It is known \cite{let,S} that every exact Courant algebroid has the following form
    \[E\cong TM\oplus T^*M,\qquad a(X+\alpha)=X,\qquad \langle X+\alpha,Y+\beta\rangle=\alpha(Y)+\beta(X),\]
    \[[X+\alpha,Y+\beta]=L_XY+(L_X\beta-i_Yd\alpha+H(X,Y,\slot)),\]
    for some closed 3-form $H$ on $M$. Two exact CAs on $M$ whose 3-forms differ by an exact 3-form can be shown to be isomorphic. In particular, all exact Courant algebroids over $M$ look locally the same.
    
    Similarly \cite{let}, every transitive Courant algebroid on $M$ is locally determined by a choice of a \emph{quadratic Lie algebra} $\mf g$ (i.e.\ a Lie algebra together with an invariant non-degenerate symmetric pairing). Explicitly, around any point in $M$ we can find an open subset $U\subset M$ and $\mf g$ such that
      \[E|_U\cong TU\oplus T^*U\oplus (\mf g\times U),\quad a(X+\alpha+s)=X,\quad \langle X+\alpha+s,Y+\beta+t\rangle=\alpha(Y)+\beta(X)+\langle s,t\rangle_{\mf g},\]
      \[[X+\alpha+s,Y+\beta+t]=L_X Y+(L_X \beta-i_Yd\alpha+\langle ds,t\rangle_{\mf g})+(L_Xt-L_Ys+[s,t]_{\mf g}).\]
      
    As a special case we can take $M=\text{point}$, so that $E=\mf g$, $a=0$, with the bracket and pairing on $E$ coinciding with those on $\mf g$. Any quadratic Lie algebra can thus be seen as a Courant algebroid. Dilatonic supergravity in this particularly simple setup is studied in detail in Subsection \ref{subsec:qla}.
  
  \section{Generalised connections and the Dirac operator}
  \label{app:connections}
    In this section we follow \cite{AX} (see also \cite{let}). On any Courant algebroid $E\to M$ we define \emph{generalised connections} to be maps satisfying
    \[D\colon \Gamma(E)\otimes \Gamma(E)\to\Gamma(E)\quad \text{s.t.\quad} D_{fu}v=fD_uv,\quad D_u(fv)=fD_uv+(a(u)f)v, \quad D_u\langle\slot,\slot\rangle=0.\]
    Here the last property again uses the fact that $D_u$ can be extended (due to the second property) to a derivation of the whole tensor algebra of $E$. Again, the definition implies that generalised connections also naturally act on spinors w.r.t.\ $E$.
    
    Slightly more surprisingly, generalised connections also naturally act on $\lambda$-densities on $M$ via
    \[D_u\sigma:=L_{a(u)}\sigma-\lambda\sigma D_\alpha u^\alpha,\qquad u\in\Gamma(E),\;\sigma\in\Gamma(L^\lambda),\]
    Thus, both $\ms L_u$ and $D_u$ act naturally on any $E$-tensors or spinors valued in $\lambda$-densities on $M$.
    
    For any generalised connection, we can define its torsion by
    \[T(u,v):=D_uv-D_vu-[u,v]+\langle Du,v\rangle.\]
    One can check that this expression is tensorial in both slots, and in fact the torsion is a tensor \[T\in \Gamma(\Lambda^3E).\]
    
    Assume now that $D$ is torsion-free (i.e.\ its torsion vanishes; such connections exist on any Courant algebroid \cite{MGF}). It then turns out \cite{AX}, crucially, that the \emph{Dirac operator} \[\slashed D:=\gamma^\alpha D_\alpha\colon \Gamma(S\otimes L^{1/2})\to \Gamma(S\otimes L^{1/2})\] is in fact independent of the choice of the particular torsion-free connection $D$ and is thus intrinsic to the Courant algebroid structure itself. Note that this independence only holds when $\di$ acts on spinor half-densities. One implication of this fact is the useful formula
    \begin{equation}\label{eq:dirac_commut}
      [\di,\ms L_u]=0.
    \end{equation}

    For instance, on an exact Courant algebroid twisted by $H\in\Omega^3_{cl}(M)$, spinor half-densities can be understood as differential forms on $M$, and the Dirac operator is
    \[\di \rho=d\rho+H\wedge \rho,\]
    while for a quadratic Lie algebra we obtain
    \[\slashed D=-\tfrac1{12}c_{\alpha\beta\gamma}\gamma^{\alpha\beta\gamma},\]
    where $c_{\alpha\beta\gamma}$ denotes the structure constants. For a transitive Courant algebroid the Dirac operator is locally a sum of the two above (where we can also take the 3-form $H$ to vanish).
    
    Another important fact about the Dirac operator acting on spinor half-densities is that its square $\di^2$ contains no derivatives and in fact corresponds to the multiplication by a function, which we will denote by \[\ms R\in C^\infty(M).\]
    As the notation suggests, $\ms R$ can be understood (up to a prefactor) as the scalar curvature associated to the generalised metric $\ms G=\on{id}$ (cf.\ the generalised Lichnerowicz formula in \cite{CSCW}).
    
    Note that \eqref{eq:dirac_commut} implies that $\ms R$ is preserved by the action of the generalised Lie derivative $\ms L$, and hence it is constant on the integral leaves of the distribution $\on{im}(a)\subset TM$. Explicitly, on a transitive Courant algebroid we obtain the constant function
    \begin{equation}\label{eq:r}
      \ms R=-\tfrac1{24} c_{\alpha\beta\gamma}c^{\alpha\beta\gamma},
    \end{equation}
    where $c$ are the structure constants of the corresponding quadratic Lie algebra $\mf g$. In particular, $\ms R$ vanishes on exact Courant algebroids.    

  \section{Some useful formulas}
    For any bosonic spinor half-densities $\alpha$, $\beta$, we have 
    \[\bar\alpha\di \beta+\bar\beta\di\alpha=\mc L_u\sigma^2,\qquad u^\alpha=\sigma^{-2}\bar\alpha\gamma^\alpha\beta\]
    for any positive half-density $\sigma$. (Note that the RHS is indeed indendent of $\sigma$ --- when multiplying $\sigma$ by any nonzero function the newly created terms involving derivatives of the function cancel against each other.) In particular
    \begin{equation}\label{eq:adj}
        \int_M \bar\alpha\di \beta=-\int_M\bar\beta\di\alpha.
    \end{equation}
    
    Let $R\to M$ be an associated vector bundle to the spin lift of the bundle of the oriented orthonormal frames of $E$ and $\lambda\in \mb R$. If $D$ is torsion free then for any $\lambda$-density valued in $R$ we have
    \[\mc L_u\psi=D_u \psi+A\cdot \psi+\lambda (D_\alpha  u^\alpha)\psi,\qquad A_{\alpha\beta}:=D_\alpha u_\beta-D_\beta u_\alpha.\]
    In particular, for spinor half-densities we have
    \[\mc L_u\psi=D_u \psi+\tfrac12(D_\alpha u_\beta)\gamma^{\alpha\beta} \psi+\tfrac12 (D_\alpha  u^\alpha)\psi.\]
    Using this, one easily shows that for any spinor half-density $\psi$ and $u\in\Gamma(E)$
    \begin{equation}\label{eq:useful}
      \slashed D(u_\alpha \gamma^\alpha \psi)=2\mc L_u\psi-u_\alpha \gamma^\alpha \slashed D\psi.
    \end{equation}

\end{document}